\documentclass[a4paper,11pt]{article}
\pdfoutput=1 % if your are submitting a pdflatex (i.e. if you have
\usepackage{jcappub} % for details on the use of the package, please
\usepackage[T1]{fontenc} % if needed

\usepackage{tikz}
\usepackage{hyperref}
\usetikzlibrary{arrows}
\usepackage{gensymb}
\title{Constraining Lorentz  Violation using 21cm and CMB Cross Correlations}

\newcommand{\red}[1]{\textcolor{red}{#1}}

\author[a]{Bhuwan Joshi,}
\author[a]{Rahul Kothari,}
\author[a]{Shyam Chaudhary}

\affiliation[a]{School of Physical Sciences, Indian Institute of Technology, Mandi, Himachal Pradesh, 175005}

\emailAdd{bhuwanj230@gmail.com}
\emailAdd{quantummechanicskothari@gmail.com}

\abstract{Lorentz symmetry is a fundamental pillar of modern Physics, yet high-energy theories often predict its violation. One potential signature of such a violation is cosmic birefringence — rotation of the polarization plane of photons due to Chern-Simons coupling in Maxwell's electrodynamics. This rotation angle, aka \textit{birefringence angle}, depends upon the distance travelled by the photon and is thus different for CMB and 21cm photons. %, we need to account for two birefringence angles $\alpha_\mathrm{CMB}$ and $\alpha_\mathrm{21cm}$. 
While the rotation angle in CMB, i.e., $\alpha_\mathrm{CMB}$, has been tightly constrained by CMB experiments, the potential of the 21cm cosmological signal to constrain this parameter, as well as constrain $\alpha_\mathrm{21cm}$, remains largely unexplored. In this work, we provide constraints on both these angles by cross-correlating 21cm and CMB signals. %We start from the modified dispersion relation due to the Chern-Simon coupling and the resulting corrections to the angular power spectra, specifically focusing on the mixing of $E$- and $B$-modes. 
Using the Fisher matrix formalism, we give our forecasts for 21cm experiments, including SKA, HIRAX, and PUMA, \& Planck like CMB experiment. %We find that the internal 21cm correlations and their cross-correlation with CMB temperature anisotropies and polarization can provide independent constraints on Lorentz violation. 
We find that best constraints $\sigma_{\alpha_\mathrm{CMB}} \sim 4.4\degree$ and $\sigma_{\alpha_\mathrm{21cm}} \sim 100\degree$ are found using $C_\ell^{T_{21} B_\mathrm{CMB}}$ and $C_\ell^{T_{21} B_{21}}$ respectively. Since birefringence hasn't yet been detected in 21cm, we choose the fiducial value $\alpha_\mathrm{21cm}^\mathrm{fid}=0$ assuming the null hypothesis.}
\begin{document}
\maketitle
\flushbottom

\section{Introduction \label{sec:intro}}
Lorentz invariance is one of the key ideas in Physics, according to which Physical laws should be consistent with Special Relativity, i.e., they should take the same form in all inertial frames that are boosted and/or rotated with respect to the frame in the picture. The standard model of particle Physics that describes three out of four fundamental interactions is based on QFTs, which are constructed to be Lorentz invariant from the beginning.% \blue{All current observations are consistent with this symmetry to a very large extent.} %The symmetry group of special relativity is Lorentz group $SO(3,1)$ consisting of %{Lorentz transformations} no matter how we move through space or time. 

However, it is crucial to investigate places where Lorentz symmetry violation may take place. %violate by employing cutting-edge observational tests. 
The fate of this symmetry at the most extreme energies of the universe remains one of the most pressing open questions in fundamental Physics. The investigation is interesting because quantum gravity theories suggest that Lorentz symmetry may not be a perfect symmetry of nature. Various approaches to quantum gravity, e.g., string theory \cite{Kostelecky:1988zi, Mavromatos:2007xe}, loop quantum gravity \cite{Gambini:1998it, Vucetich:2005ra}, spacetime noncommutativity  \cite{Carroll:2001ws}, Horava Lifschitz gravity \cite{Horava:2009uw, Wang:2017brl}, etc., often predict a minute breakdown of Lorentz invariance, typically at the Planck scale. The diffeomorphism invariance of General Relativity may be violated from spontaneous breaking of Lorentz symmetry, particularly in string theories \cite{Reyes:2022mvm, Wang:2021ctl}. Consequently, hunting for the signatures of Lorentz violation is not merely a test of a foundational principle; it is also a direct experimental pathway to probe the quantum nature of gravity at Planck scales. 

%Observations of high-energy particles from space \cite{Li:2021cdz}, like cosmic rays, gamma rays, and neutrinos, also showed some signatures of LIV. These motivated to study LIV and testing how far Lorentz invariance can really go. Within astrophysics and cosmology, the universe itself acts as a fundamental laboratory. 
A broad spectrum of approaches, spanning highly controlled laboratory experiments to observations of high-energy cosmic particles, neutrinos, gamma rays, gravitational waves, etc., \cite{Carroll:1989vb, Amram:2023jlc, Stecker:2022tzd, Li:2021cdz, AouladLafkih:2025stw, Wang:2021ctl, Gerasimov:2021chj} have been carried out. %Additionally,  high energy neutrinos aren't affected by magnetic fields, can reach from very large distances and thus can be useful for testing Lorentz symmetry \cite{Stecker:2022tzd}. 
Ref \cite{Wang:2021ctl} uses 50 gravitational wave events and analyses anisotropic birefringence, whereas  \cite{AouladLafkih:2025stw} solves the gravitational wave equation in the presence of Lorentz-violating terms. These results would be applicable in constraining symmetry violation with new gravitational wave detectors like LISA. Additionally,  Ref \cite{Gerasimov:2021chj} combines linear and circular polarization measurements of extragalactic sources for testing Lorentz violation. Interestingly, there have been suggestions to test Lorentz violation through blackhole Physics as well, e.g., by observing the shadow cast by the Kerr-like black hole with or without plasma \cite{Wang:2021irh}, analysing EHT observations under the purview of Horava gravity %model and also the EHT observations 
\cite{Liu:2025lwj}, etc.  

Theoretically, the Lorentz violation manifests in modified theories of gravity, e.g., Horava-Lifshitz, Nieh-Yan, vector-tensor, Einstein Aether, bumblebee gravity, etc., \cite{Horava:2009uw, Wang:2017brl, Cai:2021uup, Satheeshkumar:2021zvl, Ivanov:2016hng, Oost:2021tqi, Seifert:2009gi, Bluhm:2004ep, Bluhm:2007bd}. In vector tensor theories of gravity, dynamical vector fields, besides the metric, are helpful to understand spontaneous Lorentz violation \cite{Satheeshkumar:2021zvl}. %to generate massive gravitons and as models of dark matter and dark energy \cite{Satheeshkumar:2021zvl}. 
One specific case of vector tensor theory is the Einstein Aether theory, in which Lorentz invariance of General Relativity is broken by coupling a unit time-like vector field to the metric at every spacetime point \cite{Ivanov:2016hng, Oost:2021tqi}. %that has been introduced to avoid Ostrogradsky's ghosts problem and employs a Lifshitz-type anisotropic scaling between space and time, while recovering the invariance at low energies,  
Particle physicists study Lorentz violation by going beyond standard model, where Lorentz violation is characterized by tensor-valued background fields which quantify the extent of symmetry violation by modified dispersion relations \cite{Colladay:1998fq, Amram:2023jlc}. In cosmology, the generalized bumblebee model  \cite{Seifert:2009gi, Bluhm:2004ep, Bluhm:2007bd}  dynamically breaks Lorentz symmetry via a `bumblebee' vector field $(B_{\mu})$ that acquires a constant vacuum expectation value. Perturbations of this field are identified as the physical photon. A key consequence is that this ``bumblebee photon'' propagates at a speed that is direction-dependent and different from the speed of ordinary matter, leading to testable effects in resonator and cosmic-ray experiments \cite{Seifert:2009gi, Kostelecky:2003fs}.

In addition to these, a novel phenomenon occurs by introducing Chern-Simons term in Maxwell's Lagrangian \cite{Carroll:1989vb, Mewes:2008jn}. This new term, in addition to breaking Lorentz symmetry, also causes parity violation. More importantly, it changes the EM dispersion relation, thereby implying \textit{cosmic birefringence}, i.e.,  rotation of the plane of polarization of light as it travels over cosmological distances. A myriad of data sets have been employed to test the Lorentz violation signal by measuring this rotation angle. CMB experiments like WMAP, Planck, POLARBEAR, etc. \cite{Lue:1998mq, Caloni:2022kwp, Kaufman:2014rpa, Minami:2020odp, Diego-Palazuelos:2022mcp, Diego-Palazuelos:2025dmh}, have set limits on how much rotation the CMB polarization could experience. A %large number of investigations have been carried out to estimate and constrain this birefringence angle. E.g., a 
recent analysis of Planck data \cite{Minami:2020odp} reported a non-zero CMB birefringence angle $ 0.35^\circ \pm 0.14^\circ$ (68\% C.L.), (see also \cite{Diego-Palazuelos:2022mcp, Diego-Palazuelos:2025dmh}). The tightest constraints on Lorentz and CPT violation have been obtained using CMB polarization. Improvement in calibration accuracy in CMB would lead to an unambiguous measurement of parity and Lorentz violating effects \cite{Kaufman:2014rpa}. %, thereby setting indirect  limits on Lorentz violation. 
%uses Horava gravity model and also the EHT observations of supermassive blackholes to investigate signatures of Lorentz violation in real astrophysical environments. The rotation angle aka birefringence angle $\alpha$ is directly proportional to the distance travelled. 
Radio galaxies and quasar data have also been used in this regard \cite{Carroll:1989vb}. 
Furthermore, the LSS galaxy survey analysis has reported an uncertainty of $5{\degree}-15\degree$ per galaxy \cite{Yin:2024fez} {in galaxy birefringence angle}. But the potential of next-generation 21cm intensity experiments to constrain Lorentz violation remains largely unexplored. 

In this article, we constrain birefringence angles, both in CMB and 21cm signals (denoted respectively as $\alpha_\mathrm{CMB}$ and $\alpha_\mathrm{21cm}$) using their cross power spectra. We must mention that, 21cm signal has been cross correlated with other surveys, e.g., for testing local primordial non-gaussianities  \cite{Orlando:2023dgt}, %isolating cosmological signal when the foregrounds are absent using \cite{Banerjee:2022whc},
cross correlation of CMB lensing and 21cm \cite{Tanaka:2019nph}, cross correlation of 21cm and CMB polarizations \cite{Ji:2021djj} and so on. Most importantly, Ref \cite{Kadota:2019ktm} investigated cosmic birefringence by cross correlating 21cm temperature and CMB B-mode for probing Lorentz and parity violation by calculating the SNR in the reionization epochs. But to the best of our knowledge, no prior study pertaining to 21cm has constrained the birefringence angle in CMB or in 21cm. The present article serves this very purpose. From our analysis, we find that the best constraints $\sigma_{\alpha_\mathrm{CMB}} \sim 4.4\degree$ and $\sigma_{\alpha_\mathrm{21cm}} \sim 100\degree$ are found using $C_\ell^{T_{21} B_\mathrm{CMB}}$ and $C_\ell^{T_{21} B_{21}}$ respectively. Such large uncertainties on $\alpha_\mathrm{21cm}$ is due to the weak 21cm polarization signal as compared to the instrumental noise. Since 21cm polarization hasn't yet been detected, assuming the null hypothesis, we take  $\alpha_\mathrm{21cm}^\mathrm{fid} = 0$.% for providing constraints on $\alpha_\mathrm{21cm}$. 

The paper is organized as follows: in \S\ref{sec:Lorentz_violation} we introduce basic Physics of cosmic birefringence when we add the Chern-Simons term to Maxwell's Lagrangian. This leads to a modification of the photon dispersion relation, culminating in different rotations for left- and right circular polarization. This is followed by a discussion of temperature and polarization modes of both CMB and 21cm \& also their spherical harmonic coefficients in \S \ref{sec:CMB_21cm_power}. We also discuss noise power spectra for both cases. These ideas are then applied in \S \ref{sec:cross_cor_def_eq} to define cross correlations between CMB and 21cm signals. Then in \S \ref{sec:fisher_results}, we present our Fisher forecasts while also discussing the effects of foregrounds and window function on our results. We conclude in \S \ref{eq:Conclu_Outlook}. In this paper, we use fiducial $\Lambda$CDM cosmology with the best fit parameters from Planck 2018 paper \cite{Planck:2018vyg}: $h$ = 0.6766, $\Omega_\mathrm{b}h^2 = 0.02242 $ , $\Omega_\mathrm{c}h^2$ = 0.11933, $n_\mathrm{s}$ = 0.9665, $A_\mathrm{s} = 2.142\times 10^{-9}$.

\section{Modifying Maxwell Electrodynamics for testing Lorentz Violation}\label{sec:Lorentz_violation}
We study Lorentz violation by modifying the electromagnetic Lagrangian, via adding a Chern-Simons term which %This modification leads to rotation of the plane of polarization of photons as they travel from the source towards us. % One way to study LIV is by modifying the electromagnetic Lagrangian, which describes how electric and magnetic fields behave. Carroll, Field, and Jackiw’s 1990 paper\cite{Carroll:1989vb} looked into adding a Chern-Simons term to this Lagrangian. This term breaks Lorentz invariance in a way that would lead to an effect called cosmic polarization rotation (CPR) due to a phenomenon known as cosmic birefringence where the polarization (direction of oscillation) of light from faraway sources rotates as it travels to us. Observing or constraining this rotation could help us test LIV. {From their observation of distant radio galaxies, \cite{Carroll:1989vb} found that the magnitude of the vector (m = $(p_{\mu}p^{\mu})^{1/2}$) is $<1.7\times 10^{-42}h$ Gev, $h$ being the dimensionless Hubble's constant.} 
%By looking polarized light from faraway objects, like radio galaxies and quasars, to see if there’s any rotation in their polarization caused by Lorentz violation .  In this paper, we put study Lorentz violation by putting constraints on rotation angle.
%Scientists also look at polarized light from very  These measurements are sensitive to CPR and allow for tests of LIV over large cosmic distances. 
%Carroll et.al.  were the first to 
% \red{From their observation of distant radio galaxies they finds that the magnitude of the vector (m = $(p_{\mu}p^{\mu})^{1/2}$) is $<1.7\times 10^{-42}h$ Gev, where h is hubble constant.}
% \subsection{Modified Maxwell's Equations and wave solutions}
involves a coupling of an external four-vector $p^\mu$ with the dual of an electromagnetic field tensor $\tilde{F}^{\mu\nu}$. The gauge invariance then restricts the four-vector to be a constant \cite{Carroll:1989vb}.
The time component of $p^\mu$ results in Lorentz violation, whereas the spatial components define a fixed direction in space, thereby breaking rotational invariance. Maxwell's Lagrangian, after adding the Chern-Simons term, becomes
\begin{equation}
    \mathcal{L} = \mathcal{L}_\mathrm{EM} + \mathcal{L}_\mathrm{CS},\quad \mathcal{L}_\mathrm{EM}= -\frac{1}{4}F_{\mu\nu}F^{\mu\nu},\ \mathcal{L}_\mathrm{CS} = - \frac{1}{2}p_{\mu}A_{\nu}\tilde{F}^{\mu\nu} \label{eq:chern-simon-total-lagrangian}
\end{equation}
The field equations in the absence of sources, obtained after applying the Euler Lagrange equations on $\mathcal{L}$ in \eqref{eq:chern-simon-total-lagrangian} are %we get from the above Lagrangian in the absence of a source is given as
\begin{equation}
 \partial_{\alpha}F^{\alpha\beta} = p_{\mu}\tilde{F}^{\mu\beta} \label{eq:field-equation}
\end{equation}
The source term appearing in the RHS of \eqref{eq:field-equation} is due to the topological charge density sourced by the Chern-Simons term.
%We can see that under $U(1)$ gauge transformation $A_{\mu}\to A_{\mu} + \partial_{\mu}\lambda$ and using integration by parts we get \blue{[check this expression, this looks erroneous!]}
%\begin{equation}
    %\delta S_\mathrm{CS} = -\int d^4x \frac{1}{4}p_{\mu}\epsilon^{\mu\nu\rho\sigma}\partial_{\nu} F_{\rho\sigma}
    %\int d^4x\partial_{\nu}\{\frac{1}{4}p_{\mu}\epsilon^{\mu\nu\rho\sigma}\lambda F_{\rho\sigma}
%\end{equation}
%using part by integration we gets from this we can see a conserved charge which is not due to any source while due to some topological charge which we get in the right side of the maxwell's equation.
In terms of three vectors \textbf{E} and \textbf{B}, \eqref{eq:field-equation} gives the following modified forms of Gauss' and Ampere's laws (other equations remaining unchanged)
\begin{align}
    \nabla \cdot \mathbf{E} &=- \mathbf{p} \cdot \mathbf{B} \label{eq:modified-maxwell-eq1} \\
    \nabla \times \mathbf{B} &= \frac{\partial \mathbf{E}}{\partial t} -p_0 B +\mathbf{p} \times \mathbf{E} \label{eq:modified-maxwell-eq2}
\end{align}
It can be noticed that by setting the vector $p^\mu=0$, we get the standard source-free Maxwell's equations.    %no magnetic monopole and Faraday's law 
% \begin{align}
%     \nabla \cdot\mathbf{B} &= 0 \\
%     \frac{\partial\mathbf{E}}{\partial t} + \nabla\times\mathbf{B} &=0
% \end{align}
To solve the modified equations, in \eqref{eq:modified-maxwell-eq2}, we substitute the following plane wave \textit{ansatz}
\begin{equation}
    \mathbf{E}(\mathbf{r}, t) = \mathbf{E}_0 e^{i(\mathbf{k} \cdot \mathbf{r} - \omega t)},\quad
    \mathbf{B}(\mathbf{r}, t) = \mathbf{B}_0 e^{i(\mathbf{k} \cdot \mathbf{r} - \omega t)}
\end{equation}
giving the solution for the \textbf{E} field
\begin{equation}
    (\omega^2 - k^2)\mathbf{E} + \mathbf{k}(\mathbf{k\cdot\mathbf{E}}) = i (p_0\mathbf{k} - \omega\mathbf{p})\times\mathbf{E} \label{eq:equation-after-ansatz}
\end{equation}
To simplify further, we decompose the 3 vectors of the theory in the orthonormal system $\Hat{k},\Hat{e}_2$ and $\Hat{e}_3$. In this triad, the spatial component of the four vector $p^\mu=(p_0,\mathbf{p})$ can be written as $\mathbf{p} = p_{||}\hat{k} + p_2\hat{e}_2 + p_3\hat{e}_3$. In the same basis, \eqref{eq:equation-after-ansatz} % This basis allows \red{We write the four vector $p^{\mu} = (p_0,\mathbf{p})$, where } this equation can be written 
takes the following matrix form
\begin{equation}
    \sum_j M_{ij}E_j =0,\text{ where } M =
\begin{pmatrix}
\omega & -ip_3 & ip_2 \\
i\omega p_3 & \omega^2-k^2 & i(p_0k - \omega p_{||}) \\
-i\omega p_2 & -i(p_0k - \omega p_{||}) & \omega^2-k^2
\end{pmatrix}
\end{equation}
Since this is a homogeneous system of equations, for a non-zero solution for \textbf{E}, we must set $\det M=0$. Thus, we get the modified dispersion relation %after using the ansatz we get a equation in $\mathbf{E}$ as for dispersion relation relation we need to write this in the metrix form as $M_{ij}\mathbf{E}_j$. I just check we \textbf{don't} need this (0,0,k) if we use the basis  we can get det$ M_{ij}$ = 0 for any $E_{j} \neq 0$ ) which gives the following dispersion relation
%Substituting this ansatz in \eqref{eq:modified-maxwell-eq2} and choosing a coordinate system with $\mathbf{k}= (0,0,k)$ \blue{[explain why is this choice allowed?]} yields the following dispersion relation
%\begin{equation}
%(\omega^2 - k^2)\mathbf{E} + \mathbf{k}(\mathbf{k\cdot\mathbf{E}}) = i (p_0\mathbf{k} - \omega\mathbf{p})\times\mathbf{E}
%\end{equation}
%Choosing  and solving this equation gives the dispersion relation:
\begin{equation}
\omega^2 - k^2 = \pm \left( p_0 k - \omega |\mathbf{p}| \cos\theta \right)
\left[ 1 - \frac{|\mathbf{p}|^2 \sin^2\theta}{\omega^2 - k^2} \right]^{-1/2} \label{eq:dispersion-relation}
\end{equation}
where $\theta$ is the angle between the Chern-Simons vector $\mathbf{p}$ and the wave vector $\mathbf{k}$. As a simple check, if we put $p^\mu=0$, we get the standard dispersion relation back. The $\pm$ signs in \eqref{eq:dispersion-relation} correspond respectively to right and left-handed circularly polarized waves, indicating that the Chern-Simons modification to the electromagnetic Lagrangian, \eqref{eq:chern-simon-total-lagrangian} leads to \textit{birefringence}, i.e.,  different polarization states propagate with different phase velocities. Since Lorentz-violating effects are expected to be small, we only need to consider terms at linear order in the Lorentz-violating vector $p^\mu$. Thus  \eqref{eq:dispersion-relation} gives
\begin{equation}
k_\pm = \omega \mp\frac{1}{2}(p_0 - |\mathbf{p}| \cos\theta)
\end{equation}
The phase change of either the left or right circularly polarized light as it travels a distance $d$ is $\phi_\pm = k_\pm\times d$. 
The birefringence angle is then defined to be 
\begin{equation}
    \alpha = \frac{1}{2}(\phi_+ - \phi_-) = \frac{1}{2}(k_+ - k_-)d = -\frac{1}{2} (p_0 - |\mathbf{p}| \cos\theta) d
\end{equation}
The rotation angle $\alpha$ represents a measurable quantity that can be used to constrain Chern-Simons parameters $p_0$ and $\mathbf{p}$ through cosmological and astrophysical observations. We denote the birefringence angles in the case of CMB and 21cm, respectively, as $\alpha_\mathrm{CMB}$ and $\alpha_\mathrm{21cm}$. Using Fisher analysis, together with the cross-correlations of 21cm and CMB, we put constraints on both these angles. %., inherently related to Chern Simon parameter $p^\mu$ (c.f. \eqref{eq:chern-simon-total-lagrangian})  caused due to Lorentz violation.  Using Fisher analysis, we put constraint on both these parameters. %As the Lorentz violation testing will be primarily done using the otherwise absent cross correlations in case of Lorentz symmetry respect.

%\subsection{Effects of Spacetime Noncommutativity}
%There can be two separate sub-sections talking about these aspects and their consequences on the observations/Physics, etc.

%It turns out that we can observe the signatures of spacetime non-commutatiivty using TT correlations while for studying LIV we'd use cross correlations. Though we'll also assess the constraining power of cross correlations in the context of spacetime non-commutativity.

\section{CMB and 21cm Line as cosmological probes}\label{sec:CMB_21cm_power}
CMB has been an important probe for testing cosmological models. The radiation comprises photons that got decoupled out of cosmic plasma after the recombination era that happened around $z\sim 1000$. {Before decoupling, the photons suffered Thomson scattering that gave rise to linear polarization. Thomson scattering is thus the primary cause of polarization in CMB radiation, in addition to having secondary effects like Sunyaev-Zel’dovich, which happens after reionization and affects polarization only at low multipoles. As the secondary effects in CMB are subdominant by at least one order of magnitude \cite{Deutsch:2017cja}, in this study, we only focus our attention on primary effects.} {%While the CMB provides the oldest electromagnetic signal, its primary temperature fluctuations originate at very high redshift ($z \sim 1100$). {On top of CMB polarization, to effectively probe the correlation with the lower-redshift 21cm signal, we utilize CMB lensing as well.} Lensing traces the intervening matter distribution common to both 21cm and CMB photons observables \cite{Lewis:2006fu,Sarkar_2010}. %But in the case of E mode polarization, the secondary effects like Sunyaev-Zel’dovich effects at low multipole \cite{Deutsch:2017cja}, while our study focuses on the robust 21cm accessible multipoles, so we will use the primary polarization by E mode. In the new century we have 21cm intensity surveys, which can probe the universe from the dark ages to post-reionization, one of the most important epochs of the universe. In this article we will use these intensity mapping signals to constrain Lorentz violation and the parity-violating parameter, the birefringence parameter. Our study will try to focus on birefringence during the period of post-reionization, where galaxy surveys are limited due to the resolution. In this study we will focus on constraining the birefringence as a function of redshifts using the futuristic surveys SKA-MID, HIRAX, and PUMA. But the most difficult challenge of the intensity mapping surveys is the presence of foregrounds, which are an order of magnitude brighter. To mitigate the effects of foregrounds, we use the cross-correlation of the 21cm signals with the CMB. To constrain the Lorentz violation indicating parameter, we will use the Fisher matrix formalism for the 21cm surveys and future CMB surveys. In this paper, we use the cross correlation of 21cm signal from neutral hydrogen in the early universe obtained from HERA and LOFAR observatories and CMB to put constraints on the  study and set limits on Lorentz violation indicating parameter. By looking at this signal, we can check for signs of Lorentz invariance breaking over long distances in the universe. We use chern simon theory in which we introduce a Chern-Simons term in the electromagnetic Lagrangian, which adds a possible Lorentz violation effect by causing a rotation in the polarization of light as it travels. We aim to use this approach to calculate correlations of temperature with B-mode polarization, or between E-mode and B-mode polarization of power spectrum of 21cm signal and using available data from observatory like HERA and LOFAR, we try to set new limits on Lorentz violation parameters and add to the search for Lorentz violation in cosmology. 

Besides CMB, there are next-generation surveys of galaxy number counts, relying on the ability to resolve individual galaxies. This task progressively becomes harder as we go to higher redshifts. In such cases, observing the integrated emission of various lines like CO, HI, etc., is quite promising \cite{Kovetz:2017agg}. In particular, we have the 21cm line or HI that comes from neutral H, the most abundant element in the universe. The 21cm has found a lot of applications in studying cosmology --  it has been used to investigate cosmic dark ages and cosmic dawn \cite{Ali-Haimoud:2013hpa,Mondal:2023xjx,Saurabh_RRI:2021mxo}, to test general relativity and beyond  \cite{Darkmatter_21cm:2023slb,Darkmatter_21cm_2:2016sur, Hall:2012wd, Dinda:2022ixi}, testing primordial anisotropies \cite{Joshi:2025swr, Shiraishi:2016omb, Li:2019bsg}, constraining dark energy model parameters \cite{Hussain:2016fgg, Wu:2021vfz}, detectability of ultra-light axions particles \cite{Sabiu:2021aea}, constraining inflation \cite{Padmanabhan:2019xhc}, detecting relativistic dipole \cite{Lepori:2017twd} and so on. %In the present study, we use cross correlations of 21cm and CMB to constrain Lorentz invariance violation parameter $\alpha$.% Preionization 21cm signal also provides us information about the topology of the ionized region and also about clustering of collapsed halos based on the underlying matter density field \cite{Pritchard:2011xb}. 

\subsection{CMB Temperature Anisotropies and Polarization}
We give a brief summary of crucial results related to CMB (details can be found in \cite{Zaldarriaga:1996xe, Seljak:1996is}), which also serves the purpose to set the notation. %This section serves two purpose (a) setting the notation and (b) summarising some of the crucial results, that we use later . 
CMB (and also 21cm) radiation is characterized by Stokes vectors $(I, Q, U)$.  Due to Thomson scattering, circular polarization $V$ is absent \cite{Zaldarriaga:1996xe}. However, there are some non-standard interactions that can generate the same in CMB, see \cite{Bartolo:2012sd, Vahedi:2018abn, Sadegh:2017rnr}. For the purpose of this paper, we neglect the intrinsic circular polarization contribution due to these non-standard terms. For CMB, $I=\Delta T(\Hat{n})$, i.e., temperature fluctuations, is the following dimensionless quantity
\begin{equation}
    \Delta T(\Hat{n}) = \frac{T_\mathrm{CMB}(\Hat{n})}{\Bar{T}_\mathrm{CMB}} - 1
\end{equation}
where $T_\mathrm{CMB}(\hat{n})$ is the CMB temperature measured along $\hat{n}$, over-bar throughout this paper denotes averaged quantities.  %We measure temperature anisotropies  in the CMB sky using which we can define the dimensionless temperature fluctuations Components $Q$ \& $U$ represent linear polarization. % is due to Thomson scattering and is  represented by  components of the Stokes vector.%. 
Temperature fluctuations $\Delta T$ behave as a scalar under rotation in the tangent plane along the line of sight $\hat{n}$. On the contrary, linear polarization components $Q$ and $U$ transform as
\begin{equation}
    (Q\pm iU)'(\hat{n}) = e^{\mp 2i\psi} (Q\pm iU)(\hat{n})
\end{equation}
$\psi$ is the angle of rotation in the tangent plane. We thus use the $\eth$ operator \cite{Newman:1966ub} to convert $Q$ and $U$ into rotationally invariant quantities known as $E$ and $B$ modes. Then we can perform spherical harmonic expansion
\begin{equation}
    X_\mathrm{CMB}(\hat{n}) = \sum_{\ell m} X_{\mathrm{CMB},\ell m} Y_{\ell m}(\hat{n})
\end{equation}
where $X\in \{\Delta T, E, B\}$. %or the purposes of this paper, we'd be primarily interested in \red{$\Delta T$}, $E$ and $B$ modes. 
On account of mathematical properties of $\eth$ operator, $\ell\ge 2$ when $X\in\{B,E\}$. Finally, we can relate the spherical harmonic coefficients with the primordial curvature perturbations $\zeta(\mathbf{k})$ as
\begin{equation}
    X_{\mathrm{CMB},\ell m} = 4\pi i^\ell \int \frac{d^3k}{(2\pi)^3} \Delta_{X,\ell}^\mathrm{CMB}(k) Y^*_{\ell m} (\hat{k}) \zeta(\mathbf{k}) \label{eq:CMB-Harm-Coeff}
\end{equation}
where $\Delta_{X,\ell}^\mathrm{CMB}(k)$ are appropriate transfer functions calculated using CAMB software.

\subsubsection{CMB Noise Power Spectra}
We give our forecasts assuming Planck-like CMB experiments. The noise power spectrum for CMB temperature anisotropy is \cite{Smith:2006nk}
\begin{equation}N_{\ell}^{T_\mathrm{CMB} T_\mathrm{CMB}} = \left(\frac{\Delta_{T}}{\bar{T}_\mathrm{CMB}}\right)^2 \exp\left( \frac{\ell(\ell + 1)\theta_{\mathrm{FWHM}}^2}{8\ln{2}} \right)\label{eq:Noise_TT-CMB}
\end{equation}
where $\Delta_T = 1\,\mu\text{K-arcmin}$ represents the thermal noise sensitivity and $\theta_{\mathrm{FWHM}}$ denotes the full width at half maximum (FWHM) of the beam. For this analysis, we adopt  $\theta_{\mathrm{FWHM}} = 5\,\text{arcmin}$ \cite{Ji-marc:2021djj}. For the E and B modes, the noise power spectrum can be written as \cite{Gluscevic:2010vv}
\begin{equation}
    N_{\ell}^{E_\mathrm{CMB} E_\mathrm{CMB}} = N_{\ell}^{B_\mathrm{CMB} B_\mathrm{CMB}} = 2N_{\ell}^{T_\mathrm{CMB} T_\mathrm{CMB}} \label{eq:noise-cmb-EE-BB}
\end{equation}

\subsection{Brightness Temperature and Polarization of 21cm Line}
The 21cm line corresponds to the hyperfine transition between two closely spaced energy levels of the H atom known as \textit{singlet} and \textit{triplet} states. For the 21cm line,  $I $ component of Stokes vector is the excess brightness temperature relative to the CMB,  expressed as \cite{21cm_BOOK,21cm_review,Bharadwaj:2004nr}
\begin{equation}
\delta T_\mathrm{b}(\hat{n},z)= \frac{T_S- T_\mathrm{CMB}(\hat{n})}{1+z}\tilde{\tau}
\end{equation}
where the spin temperature $T_S$ decides the relative abundance of singlet and triplet states and $\tilde{\tau} \ll 1$ is the optical depth.
%These energy levels are known as .  %between hyperfine levels of the The ground state of neutral the hydrogen atom splits into the singlet and triplet states due to hyperfine transitions and the energy difference between these states corresponds to a wavelength of 21cm. \blue{some more details about Lyman-$\alpha$ etc. are also needed} The relative abundance of the singlet ($n_0$) and triplet states ($n_1$) can be written as \cite{21cm_BOOK, 21cm_review},
%\begin{equation}
   % \frac{n_1}{n_0}=\frac{g_1}{g_0}\exp{\left(-\frac{T_*}{T_S}\right)}
%\end{equation}
%where $g_1/g_0=3$ is the degeneracy of levels, $T_*=hc/{k_\mathrm{B}\lambda_{21}} = 0.068$K, is the temperature corresponding to the 21cm emission in the atom's rest frame. The spin temperature $T_S$ decides the abundance of singlet and triplet states and couples with different temperatures depending on the underlying physics \cite{Pritchard_2012}. 
%At high redshifts ($z<200$) it couples with kinetic temperature $T_K$ of the gas. At redshifts ($z<40$), due to the expansion of the universe, gas temperature density decreases, collision becomes ineffective and $T_S$ couples with the CMB temperature $T_{\gamma}$. After the formation of first luminous objects, $T_S$ couples with the Lyman $\alpha$ photons which correspond to the colour temperature $T_{\alpha}$  \cite{21cm_review,Pritchard_2012, Bharadwaj:2004nr,Ankita_Bera:2022vhw}.
Using $\delta T_\mathrm{b}(\hat{n},z)$, we can define dimensionless 21cm brightness contrast in the Fourier space as 
\begin{equation}
   \delta_\mathrm{HI}(\mathbf{k},z) \equiv \frac{\delta T_\mathrm{b}(\mathbf{k},z)}{\delta \bar{T}_\mathrm{b} (z)} - 1 = (b_{21}(z) + \mu^2f(z))\delta_\mathrm{m}(\mathbf{k},z) \label{Eq:21cm_temp_fourier}
\end{equation}
here $\mu = \Hat{k}\cdot\Hat{n}$, $\Hat{n}$ is the line of sight,  $f$ is the linear growth rate related to growth factor $D$ by $f=d\ln{D(a)}/d\ln{a}$. The first term is the $\text{HI}$ density contrast, related to dark matter density contrast $\delta_\mathrm{m}(k,z)$ through the bias parameter $b_{21}$. Since the majority of neutral hydrogen is found in self-shielded gas clouds inside galaxies in later times, the hydrogen density field is a biased indicator of the dark matter density field related through the bias. The second term corresponds to the redshift-space distortions  \cite{1987Kaiser}. We also have 21cm polarization, first studied by Babich and Loeb \cite{Babich:2005sb}. It can probe the epoch of reionization. %In this study we will only focus on the linear polarization corresponding to Stokes parameters Q and U. However the circular polarization corresponding to Stokes parameter V is subdominant across EoR.} 
The linear polarization is produced through two mechanisms: (a) intrinsic characteristics of the sources and (b) secondary processes. 
% \begin{enumerate}
%     \item Due to intrinsic characteristics of the sources: here we get the 21cm polarization by anisotropic occupation of hyperfine states.
%     % the effect is small \blue{[wrt what?]}
%     \item Secondary process: This is due to Thomson scattering of free electrons and 21cm photons. This is similar to the secondary polarization in the context of CMB that was discussed in the previous section. %In this case, radiation polarizes as it propagates towards the observer. %These are the main source of linear polarization. %comes from the secondary process, as intrinsic polarization is comparatively small. % these are similar to the secondary processes of CMB. In this case, the polarization occurs due to Thomson scattering of free electrons and the 21cm photons as discussed in the previous section.
% \end{enumerate}
%(a) intrinsic characteristics of the sources and (b) secondary processes where the radiation polarizes as it propagates towards the observer. \red{In intrinsic case, we get the 21cm polarization by a anisotropic occupation of hyperfine states, the effect is small while the secondary processes is similar to the secondary processes of CMB where the polarization occurs due to Thomson scattering of free electrons and the 21cm photons as discussed in the previous section.} {[Ref \cite{Ji:2020uro} talks about circular polarization of 21cm.]} 
In so far as the circular polarization is concerned, it arises from the interaction between the local quadrupole moment of the hydrogen atoms and anisotropies in the CMB radiation field, which induces a net spin orientation in the neutral hydrogen population. Just like CMB, we assume it to be small and thus neglect it for the purpose of our paper. %Althought they did not calculated the detectability of these signal but in reference they only mention about CMB circular polarization \cite{Mishra:2017lpz} can be detected by lunar radio base\cite{Jester_2009}} {This is should be small but in the above reference \cite{Ji:2020uro} only for the 21cm they show it for high redshifts 17,24, and 80 which is very small} %Thomson scattering of 21cm photons with the free electrons along the line of sight give us polarization and in linear polarization the Stokes parameters Q and U can be decomposed into E and B modes \cite{Seljak:1996is}. 

Using the standard line-of-sight integration method, we can get spherical harmonic coefficients %\blue{[say something about B mode of 21cm line!]} \red{The 21cm B mode can be defined in the presence of non zero primordial graviation wave ie In the presence of the tensor perturbations But the power spectrum of 21cm B mode will be small in comparion to E mode and Even thought the 21cm E mode is very small in comparion to noise power spectrum}. Just like CMB (c.f. \eqref{eq:CMB-Harm-Coeff}), we can write the harmonic coefficients 
for 21cm brightness temperature and polarization \cite{Li-Moa:2021wkb, Babich:2005sb}
\begin{equation}
   X_{{21},\ell m}(z) = 4\pi i^{\ell}\int \frac{d^3\mathbf{k}}{(2\pi)^3}\Delta_{X,\ell}^{21}(k,z)Y_{\ell m }^{*}(\Hat{k})\delta_\mathrm{m}(\mathbf{k},z) \label{eq:21cm-harmon-coeff}
\end{equation}
with the difference that in \eqref{eq:CMB-Harm-Coeff}, the harmonic coefficients didn't have any $z$ dependence. %The 21cm B mode power spectra is expected to be small as compared to noise, so for our Fisher analysis, we can safely neglect them. %The only contribution to B mode comes due to birefringence. 
In \eqref{eq:21cm-harmon-coeff}, the 21cm brightness temperature transfer functions for T and E modes are (adapted from \cite{Li-Moa:2021wkb})
\begin{align}
    \Delta_{T,\ell}^{21}(k,z) &= b_{21}(z)j_{\ell}(kr) - f(z)j_{\ell}^{\prime\prime}(kr), \\
    \Delta_{E,\ell}^{21}(k,z) &= \frac{3}{4}\sqrt{\frac{(\ell + 2)!}{(\ell - 2)!}}\Bigg[b_{21}(z)
\int_{\eta}^{\eta_0} d\eta' g(\eta^{\prime})j_2[k(\eta^{\prime} - \eta)]\frac{j_{\ell}[k(\eta_0 - \eta^{\prime})]}{[k(\eta_0 - \eta^{\prime})]^2}\notag \\ &- f(z)\int_{\eta}^{\eta_0} d\eta' g(\eta^{\prime})j_2^{\prime\prime}[k(\eta^{\prime} - \eta)]\frac{j_{\ell}[k(\eta_0 - \eta^{\prime})]}{[k(\eta_0 - \eta^{\prime})]^2}\Bigg] \label{eq:E-mode-21-transfer}
\end{align}
where again $\ell\ge 2$ for E mode in \eqref{eq:E-mode-21-transfer}. Here we should point out that auto power spectra $B_{21}B_{21}$ come in covariance along with the noise power spectrum (c.f. \eqref{eq:covariance}). Also, since $B_{21}B_{21}$ power is expected to be small as compared to instrumental noise, so we don't really need an expression for this correlation. Also, $r(z)= \eta(0) - \eta(z)$ is the conformal distance to a point at redshift $z$. %\red{Actually $\d\eta = \d r$ but the limit of integration are different for them so we have to be use different actually they are related as  = } and r is the comoving radial distance \blue{[standard symbols for comoving distance is $\eta$, have we used $s$]}. The angular power spectrum of 21cm brightness temperature contrast is give as \blue{[put some more details regarding getting this equation from 3.3]}\red{One just notation thing is that if we look at the $\delta_m(\mathbf{k},z) = T(k)D(z)\zeta(\mathbf{k})$ \cite{Babich:2005sb} were T(k) is the transfer function which is same as CMB and D(z) is the  linear theory growth function Actually CAMB gives us this products I checked it twice from CAMB and CLASS both are same.} Now we can defined the two point correlation of the spherical harmonic component of 21cm brihtness temperature, $\langle a_{\ell m}^{T}(z)a_{\ell m}^{T*}(z)\rangle = \delta_{\ell\ell^{\prime}}\delta_{mm^{\prime}}C_{\ell}^{TT}$, where $C_{\ell}$ is the 21cm TT power spectrum given as
The integration limit in \eqref{eq:E-mode-21-transfer} is from the emitted conformal time $\eta$ to the present time $\eta_0$. Also, the visibility function is defined as $g(\eta) = (d\tau/d\eta)e^{-\tau}$ and $\tau$ being the Thomson scattering optical depth \cite{Marc:2023}
\begin{equation}
    \tau (\eta) = \int_{\eta}^{\eta_0}\frac{\sigma_\mathrm{T} n_\mathrm{b} x_\mathrm{e}(\eta^{\prime})}{a(\eta^{\prime})^2} d\eta^{\prime}.
\end{equation}
Here $n_\mathrm{b}$ is the baryon comoving number density (equivalent to the present physical number density),  $\sigma_\mathrm{T}$ is the Thomson scattering total cross section and %. $2.307 \times10^{-5}(\Omega_bh^2)\text{Mpc}^{-1}$ in terms of the baryon-density parameter $\Omega_b$ is $\sigma_Tn_b$, 
$x_\mathrm{e}(\eta)$ is the mean ionisation fraction \cite{Marc:2023}
\begin{equation}
    x_e(z) = \frac{1}{2}\left\{ 1 - \tanh{\left[\frac{y(z) - y_\mathrm{re}}{\Delta_y}\right]}\right\}
\end{equation}
with $y(z) = (1 + z)^{3/2}$, $\Delta_y$ and $y_\mathrm{re}$ are the reionization model parameters.
%The cross power spectrum of 21cm brightness temperature and E-mode power spectrum $\langle a_{\ell m}^{T}(z)a_{\ell m}^{E*}(z)\rangle = \delta_{\ell\ell^{\prime}}\delta_{mm^{\prime}}C_{\ell}^{TE}$, where $C_{\ell}^{EE}$ is the 21cm EE power spectrum given as
%\begin{equation}\label{Eq:pow-TE}
%    C_{\ell}^{EE} = \frac{2}{k}\int k^2\d k\Delta_{\ell}^{T}(k,z){\Delta_{\ell}^{E}(k,z)}P_{m}(k,z)
%\end{equation}
%Equations (\ref{Eq:pow-TT}), (\ref{Eq:pow-EE}) and (\ref{Eq:pow-TE}) are respectively the TT, E-mode EE and the cross between T and E-mode angular power spectrum which we will use to probe the Lorentz violation.

\subsubsection{Instrumental Noise for 21cm Signal}
We consider radio telescopes operating in two distinct modes -- (a) single-dish (hereafter SD), in which we sum over auto correlations from all dishes and (b) interferometer (IF henceforth), where the cross correlations are combined. SD mode captures larger angular scales, whereas IF is adapted for relatively smaller ones. In this paper, we work with SKA-MID in SD and HIRAX \& PUMA in IF modes. For 21cm mapping, the dominant noise, relevant to the scales of interest, is due to the instrument, while the shot noise contribution is subdominant %For line intensity mapping, the dominant noise on the scale of our analysis is the instrument, while the shot noise contribution is subdominated
\cite{Gong:2011qf,Karagiannis:2024noise, Santos:2015noise, Bull:2014rha}. We can write the temperature noise power spectrum for both SD and IF in a single formula as \cite{Durrer:2020orn,Marc:2023}
\begin{equation}
    N_{\ell}^{T_{21} T_{21}}(z) = \frac{2\pi f_\mathrm{{sky}}T_\mathrm{sys}^2(z)}{t_\mathrm{survey}B(z)\, (\delta \bar{T}_\mathrm{b}(z) )^2}\frac{\alpha_{\ell}(z)}{\beta_{\ell}^2(z)} \label{eq:21cm-noise-pow-spec}
\end{equation}
where $T_\mathrm{sys}$ is the system temperature, $t_\mathrm{survey}$ the total observation time, the bandwidth $B(z) = \nu_{21}\Delta z/(1 + z)^2$, $\delta \bar{T}_\mathrm{b}(z)$ the mean background 21cm brightness temperature, $\alpha_{\ell}$ the dish density factor and $\beta_{\ell}$ the effective beam. Expressions for these are given in \cite{Durrer:2020orn}. %For SD and IF modes these can be written as
%\begin{equation}
   % \alpha_{\ell}^{\mathrm{SD}}(z) = \frac{1}{N_d}
%\end{equation}
%\begin{equation}
%    \beta_{\ell}^{SD}(z) = \exp{\left[-\frac{\ell(\ell + 1)}{16\ln{2}}\theta_b(z)^2\right]}
%\end{equation}
%for interferometer mode:
%\begin{equation}
 %   \alpha_{\ell}^{\mathrm{IF}}(z) = \left[\frac{\lambda(z)^2}{A_{\mathrm{eff}}}\right]^2\frac{1}{n_b(z,\ell)}
%\end{equation}
%\begin{equation}
%    \beta_{\ell}^{\mathrm{IF}}(z) = \theta_b(z)
%\end{equation}
Similar to CMB (c.f. \eqref{eq:noise-cmb-EE-BB}), we consider the noise power spectra  
\begin{equation}
    N_{\ell}^{E_{21} E_{21}} =   2N_{\ell}^{T_{21} T_{21}} \label{eq:Noise-21-EE-BB}
\end{equation}

\begin{table}
    \centering
    \begin{tabular}{l c c c}
    \hline\hline
     & HIRAX-1024 & PUMA & SKA-Mid \\
        \hline
        Redshift range & $0.775 - 2.55$ & $0.3 - 6.0$ & $0.35 - 3.05$\\
        Integration time, $t_\mathrm{survey}$ (sec) & $1.58\times 10^8$ & $1.58\times 10^8$ & $3.6\times 10^7$ \\
        Sky fraction, $f_\mathrm{sky}$ & $0.36$ & $0.5$  & $0.49$ \\
        Dish diameter, $D_\mathrm{dish}$ (m) & 6 & 6 & 15 \\
        Maximum baseline & 0.25 km & 1.0 km & 150 km \\
        $N_\mathrm{dish}$ & 1,024 & 32,000 & 197 \\
        \hline\hline
    \end{tabular}
    \caption{The instrumental details of the HIRAX-1024 and PUMA radio interferometers used in this work \cite{CosmicVisions21cm:2018rfq,Karagiannis:2022ylq}. Information about SKA-Mid (Band 1) is based on data from the official website \href{https://www.SKA-Mid.int/en/explore/telescopes/ska-mid}{SKA-MID}.}
    \label{table:Radio_arry}
\end{table}

\section{Cross Correlations for studying Lorentz Violation} \label{sec:cross_cor_def_eq}
The spherical harmonic coefficients of $T$, $E$ and $B$ modes, under parity  transform as 
\begin{equation}
    T_{\ell m} \longrightarrow (-1)^\ell T_{\ell m},\ E_{\ell m} \longrightarrow (-1)^\ell E_{\ell m},\ B_{\ell m} \longrightarrow (-1)^{\ell +1} B_{\ell m}
\end{equation}
These properties hold for both CMB and 21cm signals. By ensemble average of spherical harmonic coefficients, we can define the angular power spectra
\begin{equation}
    \langle X_{\ell m}^a (Y_{\ell'm'}^b)^*\rangle = \delta_{\ell \ell'} \delta_{mm'} C_\ell^{X_aY_b} \label{eq:pow-spec-def-no-biref}
\end{equation}
where $X,\ Y\in  \{\Delta T,E,B\}$ denote the mode being considered and $a,\ b\in\{\mathrm{CMB}, \mathrm{21cm}\}$, the corresponding signal. The harmonic coefficients can have redshift dependences (e.g., for the 21cm case) that we don't show explicitly for notational brevity. The Kronecker deltas arise from statistical isotropy. As the power spectra is parity invariant, it follows that the cross correlation of B mode of either CMB or 21cm with the remaining fields is zero, i.e, $C_\ell^{X_aB_b}=0$ for $X\ne B$, e.g., $C_\ell^{T_{21}B_{21}}= C_\ell^{E_\mathrm{CMB}B_{21}} =0$, etc. However, this is no longer true in the presence of birefringence, where mixing of $E$ and $B$ modes takes place, and thus we can write  \cite{Komatsu:2022nvu} 
\begin{align}
T^\mathrm{O}_{a,\ell m} & = T_{a,\ell m} \\
    {E}_{a,\ell m}^\mathrm{O} &= E_{a,\ell m}\cos(2\alpha_a) - B_{a,\ell m}\sin(2\alpha_a) \label{eq:E-transform-biref} \\
    {B}_{a,\ell m}^\mathrm{O} &= E_{a,\ell m}\sin(2\alpha_a) + B_{a,\ell m}\cos(2\alpha_a) \label{eq:B-transform-biref}
\end{align}
where again $a\in\{\mathrm{CMB},\ \mathrm{21cm}\}$ and ``O'' in the superscript denotes the harmonic coefficients in the presence of birefringence. Thus, in the presence of birefringence, $E$ and $B$ modes are mixed; however, the $T$ mode remains unaffected. This is true for both 21cm and CMB signals. Clearly, $\alpha_\mathrm{CMB} \ne \alpha_\mathrm{21cm}$ as the photons in the two cases will travel different distances and thus will have different birefringence angles. In the presence of birefringence, we define the cross power spectra just like in \eqref{eq:pow-spec-def-no-biref} as
\begin{equation}
    \langle X_{a,\ell m}^\mathrm{O} (Y_{b,\ell'm'}^\mathrm{O})^*\rangle = \delta_{\ell \ell'} \delta_{mm'} C_\ell^{X_aY_b,\mathrm{O}} \label{eq:pow-spec-def-biref}
\end{equation}
Using \eqref{eq:E-transform-biref} and \eqref{eq:B-transform-biref}, together with the definition \eqref{eq:pow-spec-def-biref}, it follows that
\begin{align}
    C_{\ell}^{T_aB_b,\mathrm{O}} &= C_{\ell}^{T_aE_b}\sin(2\alpha_b) \label{Eq:TB:birefri} \\
    C_{\ell}^{E_aB_b,\mathrm{O}} &= C_{\ell}^{E_aE_b}\sin{(2\alpha_b)}\cos{(2\alpha_a)} - C_{\ell}^{B_aB_b}\sin{(2\alpha_a)}\cos{(2\alpha_b)}\label{Eq:EB:birefri}
\end{align}
We can simplify these expressions by noting that (a) the birefringence angles are small, i.e. $\alpha_a\ll 1$, and (b) B-mode power spectra are small as compared with E-mode, which gives 
% \begin{align}
%     C_{\ell}^{T_aB_b,\mathrm{O}} &= C_{\ell}^{T_aE_b}(2\alpha_b) \label{Eq:TB:birefri_NEW} \\
%     C_{\ell}^{E_aB_b,\mathrm{O}} &= C_{\ell}^{E_aE_b}{(2\alpha_b)} - C_{\ell}^{B_aB_b}(2\alpha_a) \label{Eq:EB:birefri_NEW}
% \end{align}
% Now using the fact that B mode is smaller than E mode we can neglect the second term of eq. \ref{Eq:EB:birefri_NEW} here so our final equations will look like
\begin{align}
    C_{\ell}^{T_aB_b,\mathrm{O}} &= 2\alpha_b\ C_{\ell}^{T_aE_b} \label{Eq:TB:birefri_final} \\
    C_{\ell}^{E_aB_b,\mathrm{O}} &= 2\alpha_b\ C_{\ell}^{E_aE_b} \label{Eq:EB:birefri_final}
\end{align}
From \eqref{Eq:TB:birefri_final} and \eqref{Eq:EB:birefri_final}, it is clear that the angles $\alpha_\mathrm{CMB}$ and $\alpha_\mathrm{21cm}$ don't couple and thus we can constrain them separately. This means our Fisher Matrix takes a diagonal form. For our analysis, we consider the fiducial value of the CMB birefringence angle
%The best part of our last equation \ref{Eq:EB:birefri_final} is that there is not mixing of different $\alpha's$ which is present in the quation \ref{Eq:EB:birefri_NEW}, so we don't need the fisher matrix our ongoing work is fine with one change that is the 
$\alpha_\mathrm{CMB}^\mathrm{fid}=0.3\degree$ as has been used by Planck \cite{Diego-Palazuelos:2022mcp}. As the birefringence hasn't yet been detected in 21cm radiation, we constrain $\alpha_\mathrm{21cm}$ adopting the null hypothesis and consider $\alpha_\mathrm{21cm}^\mathrm{fid}=0$.
%of birefringece anlge is $\alpha_{\mathrm{CMB}} = 0.3 $ while for 21cm we use the null hypothesis were the $\alpha_{21}$ for the covariance is zero and in derivative part the $\alpha_{21}$ is not present due to differentiation. We use these equations to constrain Lorentz violating parameter \red{parameters $\alpha_{\mathrm{CMB}}$ and $\alpha_{21}$} $\alpha$ in the next section. 

We now enumerate all possible cross-correlations between CMB and 21cm signals that can be used to constrain {$\alpha_a$}. Since one of the fields in the cross correlation has to be the B mode (either from 21cm or CMB), we have 8 possibilities -- out of these, $ E_\mathrm{CMB} B_\mathrm{CMB} $ and $ T_\mathrm{CMB} B_\mathrm{CMB} $  have already been considered for constraining {$\alpha_{\mathrm{CMB}}$} by Planck \cite{Planck:2016soo}. So we give our constraints using -- (a) $ T_\mathrm{21} B_\mathrm{21} $ (b) $ E_\mathrm{21} B_\mathrm{21}$ (c) $ E_\mathrm{CMB} B_\mathrm{21}$ (d) $ T_\mathrm{CMB} B_\mathrm{21}$ (e) $E_\mathrm{21} B_\mathrm{CMB}$ and (f) $T_{21} B_\mathrm{CMB}$. Thus first four correlations are used to put constraints on $\alpha_\mathrm{21cm}$ while the last two are used to constrain $\alpha_\mathrm{CMB}$.

\section{Fisher Forecasts}\label{sec:fisher_results}
%In this section, we use the Fisher  formalism to constrain birefringence angles  $\alpha_{\mathrm{CMB}}$ and $\alpha_{21}$. 
We define our Fisher matrix as   \cite{Gluscevic:2010vv}
% \blue{The most general fisher matrix with the only assumption of guassianity is \begin{equation}
%     F_{\alpha\beta} = \frac{1}{2}Tr[C^{-1}C,{\alpha}C^{-1}C,\beta + 2C^{-1}\mu,\alpha\mu^{t},\beta]
% \end{equation} If we have the gaussian data $\mathbf{x}$ with mean $\mu$ and covariance matrix C is defined in terms of data matrix D = $(\mathbf{x} - \mathbf{\mu})(\mathbf{x-\mu})^{t}$ as
% \begin{equation}
%     \langle D \rangle = C
% \end{equation}}
\begin{equation}\label{eq:Fisher_mat_main}
F^{X_aY_b}_{\alpha_a \alpha_b}(z) = \sum_{\ell = \ell_{\min}}^{\ell_{\max}} 
\frac{\partial C_{\ell}^{X_aY_b,\mathrm{O}}}{\partial \alpha_a}\,\frac{\partial C_{\ell}^{X_aY_b,\mathrm{O}}}{\partial \alpha_b}
% \frac{\partial C_{\ell}^{XY,\mathrm{O}}}{\partial p_j} 
\frac{1}{\mathrm{Cov}(\tilde{C}_{\ell}^{X_aY_b,\mathrm{O}}, \tilde{C}_{\ell}^{X_aY_b,\mathrm{O}})}
\end{equation}
where $F^{X_aY_b}_{\alpha_a \alpha_b}$ is the Fisher matrix element obtained using the cross correlation $C_\ell^{X_aY_b}$. %As was discussed in the previous section, all cross correlations to be considered in this paper have been enumerated in the previous section. As was discussed in the previous section, the two parameters aren't coupled so we can constrain them separately. This is equivalent to the Fisher matrix being diagonal.  
The covariance, on account of the fact that birefringence angles in \eqref{eq:E-transform-biref} \& \eqref{eq:B-transform-biref} are small, and thus we need to keep only $\mathcal{O}(\alpha_a)$ terms, can be written as
\begin{equation}\label{eq:covariance}
    \mathrm{Cov}(\tilde{C}_{\ell}^{X_aY_b, \mathrm{O}}, \tilde{C}_{\ell}^{X_aY_b, \mathrm{O}}) \approx \frac{1}{(2\ell + 1)f_\mathrm{sky}} \left(\tilde{C}_{\ell}^{X_aX_a}\tilde{C}_{\ell}^{Y_bY_b} + \left(\tilde{C}_{\ell}^{X_aY_b,\mathrm{O}}\right)^2\right) 
\end{equation}
where quantity with tilde represents the noise added angular power spectrum. Following \cite{Gluscevic:2010vv}, we consider a zero noise power spectrum for cross-correlations between different fields and signals. For auto correlation, the expressions for noise power spectra $N_\ell^{X_\mathrm{CMB}X_\mathrm{CMB}}$ with $X\in \{\Delta T, E, B\}$ are given in \eqref{eq:Noise_TT-CMB} and \eqref{eq:noise-cmb-EE-BB} while that for 21cm in  \eqref{eq:21cm-noise-pow-spec} and \eqref{eq:Noise-21-EE-BB}.
%\blue{From the above reference \cite{Gluscevic:2010vv} of eq.9. The observable quantity %from the observed map is $d_{\ell m}^{X}$ whose ensemble average gives the $\langle(d_{\ell m}^X)d_{\ell m}^Y\rangle = C_{\ell}^{XY} + N_{\ell}^{XY}$ where $C_{\ell m}^{XY}$ is the angular power spectrum theoretical and $N_{\ell}^{XY}$ is the noise power spectrum. Now define the estimator for the angular power spectrum as \begin{equation}
%   \hat{C}_{\ell}^{XY} = \sum_{m = -\ell}^{\ell}\frac{d_{(\ell m}^X)^*d_{\ell m}^Y}{(2\ell+1)} - N_{\ell}^{XY} \end{equation}
%Then the power spectrum covariance matrix is given as 
%\begin{equation}
 %   \mathrm{Cov}_{\ell}^{XYWZ} = \langle(\hat{C}_{\ell}^{XY} - C_{\ell}^{XY})(\hat{C}_{\ell}^{WZ} - C_{\ell}^{WZ})\rangle
%\end{equation}
%The solution of above equation gives the equation \ref{eq:covariance}.} = C_{\ell}^{X_aY_b, \mathrm{O}} + N_{\ell}^{X_aY_b}
Finally, $\ell$ in \eqref{eq:Fisher_mat_main} ranges over values that depend upon both (a) redshift $z$ and (b) the survey mode being employed. %over $\ell$ values is taken from the accessible angular scales of the surveys. 
The accessible multipoles for single-dish (SD) and interferometer (IF) modes are \cite{Kothari:2023keh}
\begin{equation}
\ell_\mathrm{min}(z) = 
\begin{cases}
    2 & \mathrm{SD} \\ {150}/{(1+z)} & \mathrm{IF}
\end{cases},\quad \ell_\mathrm{max}(z) = \begin{cases}
    {450}/{(1+z)} & \mathrm{SD} \\ 500 & \mathrm{IF} \label{eq:multipoles-IF-SD}
\end{cases}
\end{equation}
%The covariance where $C^{XY,\mathrm{O}}$ is the angular power spectrum  with the cosmic birefringence, and the covariance $\mathrm{Cov}(C_{\ell}^{XY}, C_{\ell}^{WZ})$ can be written as
% \begin{equation}
%     \mathrm{Cov}(\hat{C}_{\ell}^{XY}, \hat{C}_{\ell}^{WZ}) = \frac{1}{(2\ell + 1)\mathrm{f_{sky}}} \left\{\hat{C}_{\ell}^{XW}\hat{C}_{\ell}^{YZ} + \hat{C}_{\ell}^{XZ}\hat{C}_{\ell}^{YW}\right\}
% \end{equation} In this analysis, we assume vanishing noise cross-correlations following . 
Leveraging the tomographic nature of intensity mapping experiments, we compute the total Fisher information by summing over independent redshift bins, $F_{\alpha_a \alpha_b,\mathrm{tot}}^{X_aY_b} = \sum_{z_i}F^{X_aY_b}_{\alpha_a \alpha_b}(z_i)$ in the corresponding redshift range. Since the Fisher matrix is diagonal, we define 
\begin{equation}
    \sigma_{\alpha_a} = \Big(F_{\alpha_a \alpha_a,\mathrm{tot}}^{X_bY_a}\Big)^{-1/2}
\end{equation}
using cross correlation $C_\ell^{X_bY_a}$. Following \cite{Karagiannis:2022ylq, Kopana:2024qqq, Rossiter:2024tvi}, we use redshift bin width $\Delta z = 0.1$. Also, as was discussed in the previous section, we use $\alpha_\mathrm{CMB}^\mathrm{fid}= 0.3\degree$ and $\alpha_\mathrm{21cm}^\mathrm{fid}=0$

%\red{We are using both initially We had the problem with the BB mode, is very small and less explored in 21cm signals but as the B mode information only comes in covarinace with the noise and in case of 21cm noise (instrumental) is order of magnitude greater than the B mode so we don't need the expression we can just add the Noise only. Lilerature suggests that the EB equation is good due to its low SNR but studing both will be helpfull} We expect that the auto correlation $T^{21}T^{21}$ will be orders of magnitude larger than the cross correlation $T^\mathrm{CMB}E^{21}$.

\begin{figure}[t]
\centering
\includegraphics[width=0.45\textwidth]{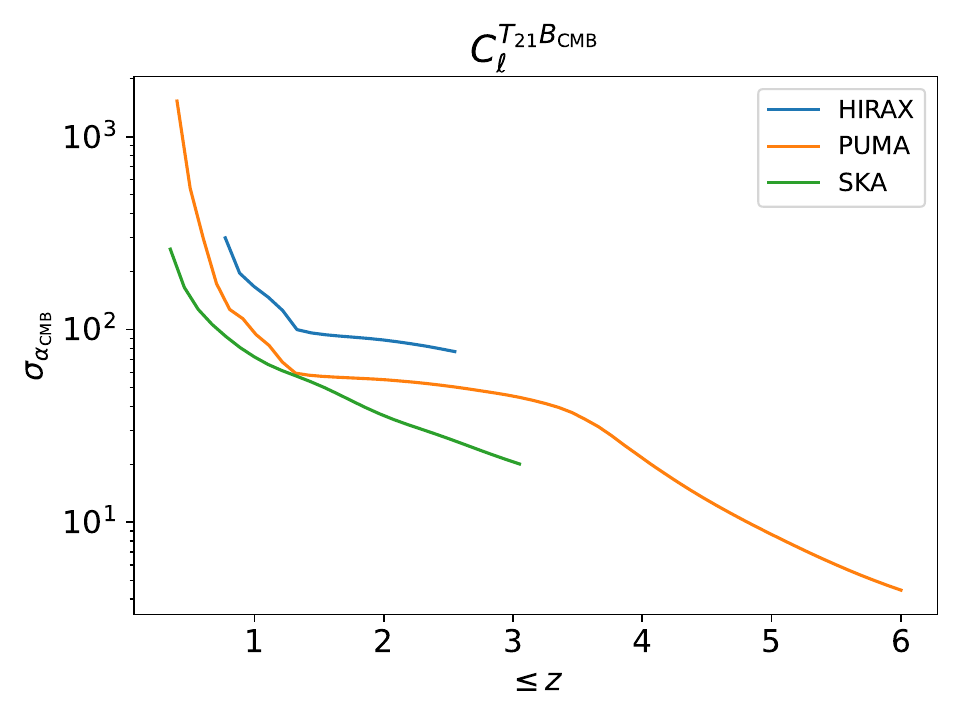}
\includegraphics[width =0.45\textwidth]{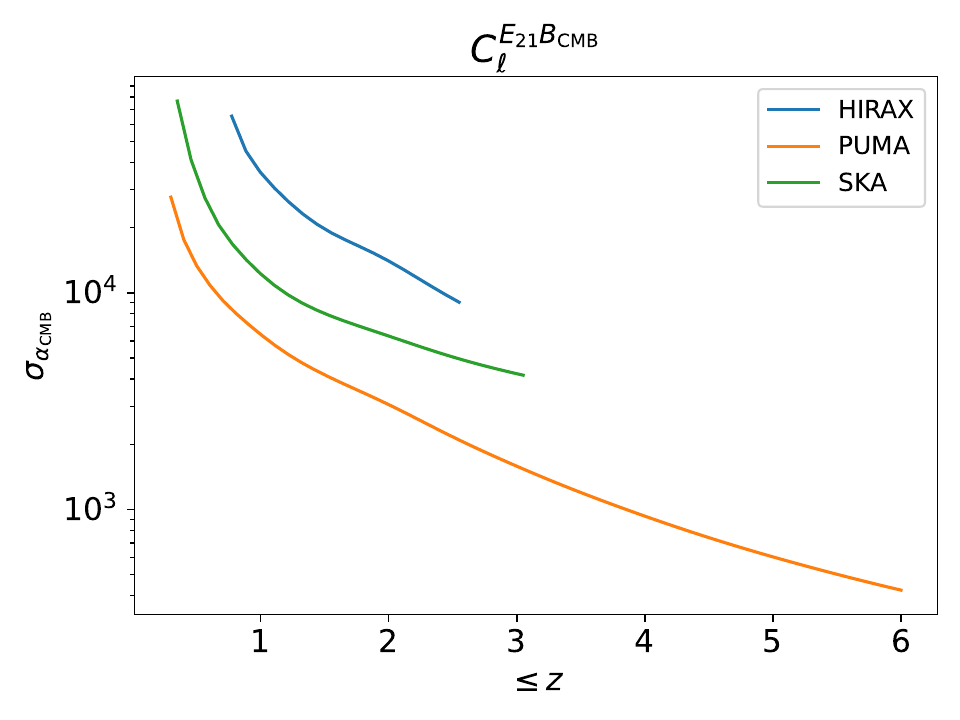}
\caption{Cumulative error on the birefringence angle $\alpha_\mathrm{CMB}$ (in degrees) as a function of redshift using various correlations. In these plots, we have used $\alpha_\mathrm{CMB}^\mathrm{fid} = 0.3\degree$. The bin-width for the Dirac window function is chosen to be $\Delta z= 0.1$. The correlations considered are \textit{left:}  $C_{\ell}^{T_{21} B_{\mathrm{CMB}}}$ and \textit{right:}  $C_{\ell}^{E_{21} B_{\mathrm{CMB}}}$. It is clear that $C_{\ell}^{T_{21} B_{\mathrm{CMB}}}$ gives the best constraint on $\alpha_\mathrm{CMB}$.}
\label{fig:fisher-constraints}
\end{figure}

\subsection{Effects of 21cm Foregrounds and Window Function}
Before presenting our results in the next section, we'd like to discuss the effects of foregrounds and window functions on our constraints. It is well known that foreground levels in 21cm are 5 orders of magnitude higher than the signal \cite{Alonso:2014dhk, Spinelli:2021emp, Shaw:2014khi}, thereby playing an important role in auto power and bispectrum estimation. It is also a very well-known fact that contaminants of individual surveys are biased, thus they don't {affect} cross correlation  (see \cite{Padmanabhan:2019xhc, Oxholm:2021zxp, Chen:2022itx} and references therein). So it might appear that foregrounds play no role in covariance for cross-correlation. However, from \eqref{eq:covariance}, it is clear that even for cross correlations, the covariance term contains auto power spectra like $C_\ell ^{T_{21}T_{21}}$ and $C_\ell^{E_{21}E_{21}}$. Now we assess the effect of foregrounds on our constraints. Since foregrounds contaminate only large angular scales, their effects are considered by the cut $\ell_\mathrm{min}^\mathrm
{fg}\ge 5$ \cite{Durrer:2020orn}. However, from \eqref{eq:multipoles-IF-SD}, we find that $\ell_\mathrm{min} \ge 20$ for all the redshifts considered in this paper. Thus, we conclude that foreground effects on $C_\ell ^{T_{21}T_{21}}$ don't alter our constraints for surveys in IF modes (PUMA and HIRAX). But they definitely play an important role, albeit very small, when we consider surveys in SD mode (SKA-Mid). %for IF mode is always greater than this value. Thus we are already including foreground effects in our analysis for $C_\ell ^{T_{21}T_{21}}$. 
%Foreground effects on angular power spectra has been estimated in \cite{Kothari:2023keh}. In order to estimate the effects of foregrounds on constraints, we take the maximum ratio of power spectra for either SD or IF mode, taken over $\ell$ and redshifts $z$
% \begin{equation}
%     R = \max_{\ell,z} \frac{C_\ell^{T_{21}T_{21}}}{C_\ell^{T_{21}T_{21},\mathrm{fg}}}
% \end{equation}
% where the power spectra in the denominator includes the foreground effects \cite{Kothari:2023keh}. We then divide the power spectra in \eqref{eq:covariance} to account for the foregrounds. We find that there isn't significant change in the constraints in the two cases. E.g., for the cross correlation $C_{\ell}^{T_{21}B_{\mathrm{CMB}}}$ as $\sigma_{\alpha_{\mathrm {CMB}}} = 3.74\degree$ in the presence of foregrounds, while in the case of of without foregrounds \sigma_{\alpha_{\mathrm {CMB}}} = 4.44\degree$. This suggests that the effect of foregrounds in constraining the parameters is quite small and thus for our subsequent analysis, we give these constraints without consider foregrounds due to numerical complexities. 
For 21cm polarization, %, a more careful treatment is needed to account for foreground and other effects \cite{Bernardi:2013dra, WMAP:2006rnx, Lenc_2016, Asad_2018, Jelic:2010vg, Moore:2013ip}. Nonetheless, 
the instrumental noise is orders of magnitude higher than $C_\ell^{E_{21}E_{21}}$. We expect that since foregrounds will only reduce this signal even further, there will be no effect of foregrounds in this case, too. We give our constraints considering all these facts into consideration. %Thus in this paper, we neglect foreground effects on $C_\ell^{E_{21}E_{21}}$ that might alter our constraints and leave a more detailed assessment for future work.

We now turn our attention to the window function. Although the tophat function is more realistic for making comparisons with the observations, we found that using a Dirac window in place of tophat doesn't change the results significantly. E.g.,  considering PUMA  and $C^{T_{21}B_\mathrm{CMB}}_\ell$, we find {$\sigma_{\alpha_{\mathrm{CMB}}}$} $3.18^\circ$ and $4.42^\circ$ respectively for tophat and Dirac windows. Thus, to avoid numerical complications, for all subsequent analysis, we use the Dirac window function.

\subsection{Results and Discussion \label{sec:res_method}}
%the fiducial value of birefringence angle as $0.3\degree$ for both 21cm and CMB \cite{Planck:2016soo, Komatsu:2022nvu, Diego-Palazuelos:2022mcp}. \red{The fiducial value of the birefringence angles $\alpha_{\mathrm{CMB}} = 0.3$ and for the 21cm $\alpha_{21}$ we use the null hypothesis.}
% Before we present our numerical results, we discuss assumptions that underlie our analysis
% \begin{itemize}
% \item We  
% \item We should also mention that since we are considering cross correlations between various fields and probes, this will mitigate the effects of foregrounds which are very prominent when we consider auto correlations of 21cm \cite{Alvarez:2005sa} as was discussed at the end of \S\ref{sec:fisher_results}.
% \item We 
% \end{itemize}
We present the Fisher matrix forecasts using various experiments -- SKA, HIRAX, and PUMA. %for  and $\alpha_\mathrm{21cm}$ respectively in Figures  and \ref{fig:fisher-constraints-21cm}. %(see Table \ref{table:Radio_arry}). % calculated using Eq. \ref{eq:Fisher_mat_main}. We analyze the 21cm signal (Eq. \ref{eq:21cm-harmon-coeff}) and its cross-correlation with the Cosmic Microwave Background (CMB). For the CMB analysis, we include CMB lensing as well as E- and B-mode polarization, following Eqs. \ref{Eq:lensing_harmo} and \ref{eq:CMB-Harm-Coeff}. We apply the Fisher matrix formalism to various 21cm surveys, specifically . For the CMB, we incorporate instrumental noise as defined in subsection \ref{subsec:Noise}. As per the discussion done before, we present forecasts for all 6 correlations for in Figure . In the figure, we see following features
\begin{enumerate}
\item In Figure \ref{fig:fisher-constraints}, we give our constraints for $\alpha_\mathrm{CMB}$. The best constraints $\sigma_{\alpha_\mathrm{CMB}} \sim 4\degree$ on $\alpha_\mathrm{CMB}$ are obtained using $C_{\ell}^{T_{21} B_{\mathrm{CMB}}}$. This is due to the fact that the 21cm brightness temperature signal is larger as compared with $E_\mathrm{21}$. This results in large covariance and hence better constraints. % and CMB polarization maps have high . %\red{ie the instrumental noise level is less than the signal}. The other combination $C_\ell^{T_{21}B_{21}}$ yield relatively poor constraints because the theoretical signal is orders of magnitude below the instrumental noise levels, resulting in a large covariance \red{\ref{eq:covariance} first term} that drastically increase the parameter uncertainties. 
\item Constraints provided by $C_{\ell}^{T_{21} B_{\mathrm{CMB}}}$ are less competitive than the CMB-only limits, e.g., \cite{Diego-Palazuelos:2022mcp, Diego-Palazuelos:2025dmh}. % discussed in Section~\ref{sec:intro}. 
This can be attributed to the weak correlation between 21cm and CMB signals, which further happens due to the large redshift separation between the primary CMB from the surface of last scattering and the 21cm signal in the post-reionisation epoch.% " In this analysis, we cross-correlate the 21cm intensity field ($T_{21}$) with the CMB primary E-modes ($E_{\mathrm{CMB}}$). However, this correlation is weak 

\item In Figure \ref{fig:fisher-constraints-21cm}, we give constraints on $\alpha_\mathrm{21cm}$. The best constraints $\sigma_{\alpha_\mathrm{21cm}}\sim 100 \degree$ are obtained using $C_{\ell}^{T_{21}B_{21}}$. %due to the strong correlation between 21cm T and E in comparison to CMB. One of the reasons for getting no so good constraints is the fact that cross correlation part in \eqref{eq:covariance} won't contribute when we consider fiducial value $0$. Mathematically, the inclusion of this component increases the total covariance, thereby weakening the resulting constraints. 
Such a poor constraint is due to a weak 21cm polarization signal as compared to the instrumental noise. %Furthermore, the 21cm polarization signal remains difficult to probe because the instrumental noise significantly exceeds the theoretical signal. 
We are optimistic that in the near future, once next-generation instrumentation reduces the noise levels even further, more stringent constraints will become achievable. Future experiments achieving significantly lower thermal noise levels may unlock the potential of the 21cm E-mode. %Since $E_{21}$ and $T_{21}$ originate from the same redshift, their auto and cross correlations would yielding far tighter constraints on birefringence once the signal becomes distinguishable from instrumental noise.
%\item In some cases IF mode and in some cases SD more works better also in some cases SKA and in some cases (for a given redshift) PUMA works better? {The possible reason for such behaviour is the trade between the linear perturbation theory and the noise levels of signals in case of $C_{\ell}^{T_{21} B_{\mathrm{CMB}}}$ SKA works best for its redshifts range because of linear theory theory which restrict the k or equivalently the $\ell$ so more information ell will be small ell values (meaning power spectrum becomes small or decays after some ell values) in which the SKA or single dish works good while for the other involved the 21cm E mode which are very small in comparison to noise so the noise decides the properties and the noise of PUMA has the small noise so which shows the best constraint}
%\item In other correlations, the error obtained is larger, so the cumulative error obtained after considering all the cross correlations won't change much. 
\end{enumerate}

%The resulting error forecasts are presented in three figures. Figure \ref{fig:T_21_B_CMB} illustrates the forecasted errors for the cross-correlations: the left panel shows the 21cm brightness temperature with CMB B-modes, while the right panel displays the 21cm E-mode polarization with CMB B-modes. Similarly, Figure \ref{fig:T_21B_21} displays the forecasted errors for the internal 21cm correlations: the left panel presents the 21cm brightness temperature with 21cm B-modes, while the right panel shows 21cm E-modes with 21cm B-modes. Finally, Figure \ref{fig:T_CMB_B_21} illustrates the cross-correlations between the CMB and 21cm B-modes; the left panel shows CMB lensing cross-correlated with 21cm B-modes, and the right panel depicts CMB E-modes with 21cm B-modes.

\begin{figure}[t]
\centering
\includegraphics[width=0.45\textwidth]{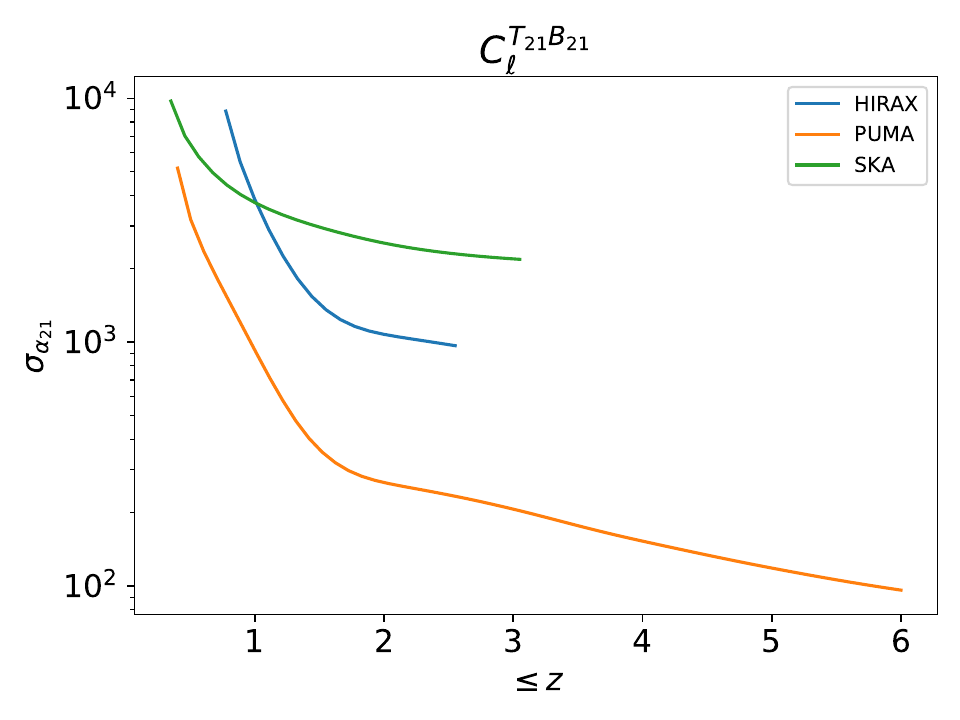}
\includegraphics[width =0.45\textwidth]{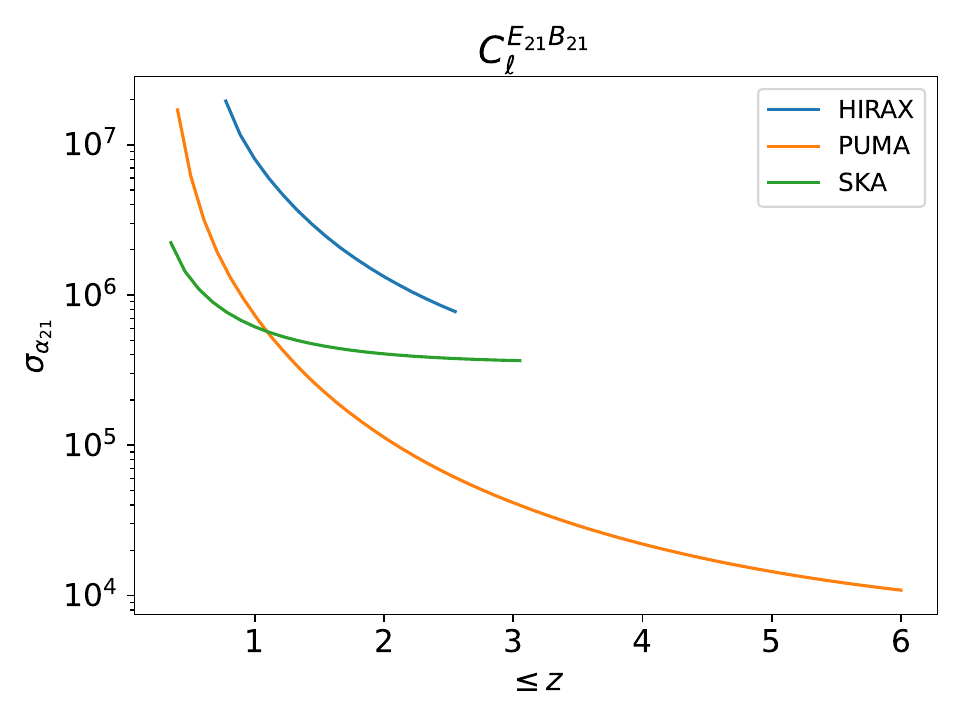}
\includegraphics[width=0.45\textwidth]{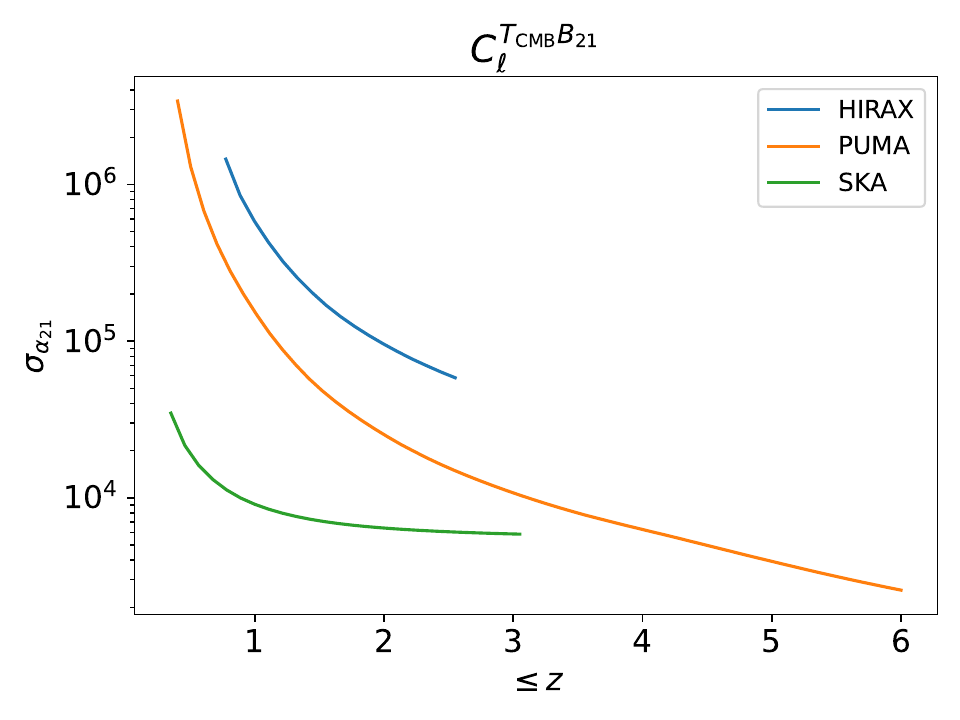}
\includegraphics[width =0.45\textwidth]{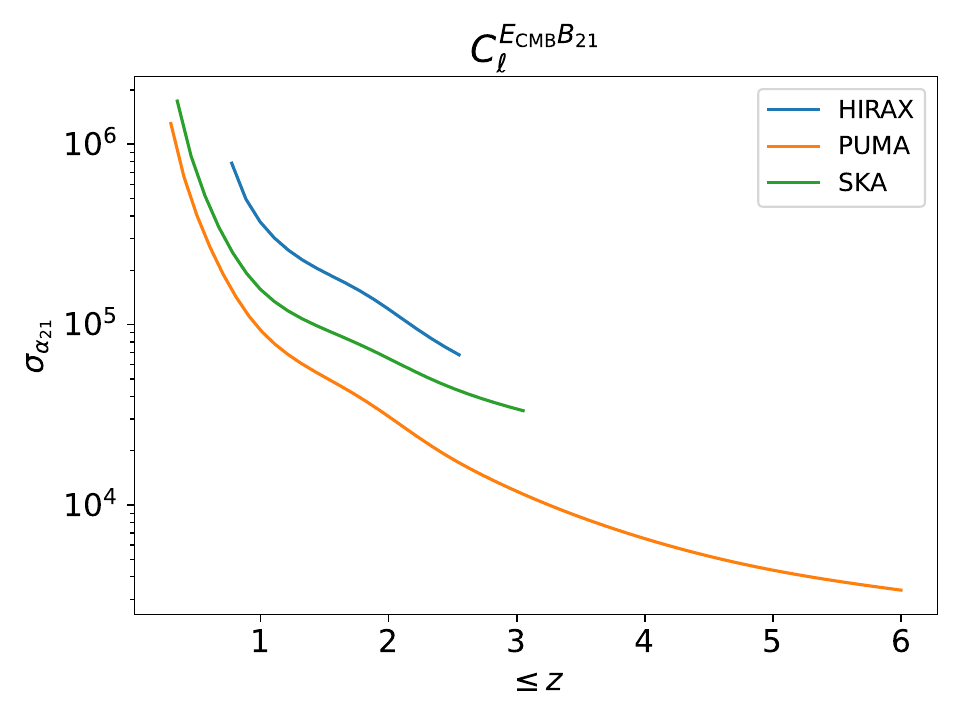}
\caption{Cumulative error on birefringence angle $\alpha_\mathrm{21cm}$ (in degrees) as a function of redshift using various correlations. Here, we've used  $\alpha_\mathrm{21cm}^\mathrm{fid} = 0$  (for discussion, see \S \ref{sec:cross_cor_def_eq}). \textit{Row 1:} left $C_{\ell}^{T_{21}B_{21}}$ and \textit{right:} $C_{\ell}^{E_{21}B_{21}}$, \textit{Row 2:} \textit{left:} $C_{\ell}^{T_{\mathrm{CMB}} B_{21}}$ and \textit{right:}  $C_{\ell}^{E_{\mathrm{CMB}} B_{21}}$. It is clear that $C_{\ell}^{T_{21}B_{21}}$ has the best constraining power out of all possibilities.}
\label{fig:fisher-constraints-21cm}
\end{figure}

%\begin{figure}[t]
%\centering
%\includegraphics[width=0.4\textwidth]{figures/Sigma_val_T_21B_21.pdf}
%\includegraphics[width =0.4\textwidth]{figures/Sigma_val_E_21_B_21.pdf}
%\caption{Cumulative error on the rotational angle $\alpha$ as a function of redshift using the 21cm cross correlation with 21. Left is using, $C_{\ell}^{T_{21}B_{21}}$ right is using $C_{\ell}^{E_{21}B_{21}}$.}
%\label{fig:T_21B_21}
%\end{figure}

%\begin{figure}[t]
%\centering
%\includegraphics[width=0.4\textwidth]{figures/Sigma_val_T_CMB_B_21.pdf}
%\includegraphics[width =0.4\textwidth]{figures/Sigma_val_E_CMB_B_21.pdf}
%\caption{Cumulative error on the rotational angle $\alpha$ as a function of redshift using the 21cm cross correlation with 21. Left is using, $C_{\ell}^{T_{\mathrm{CMB}}B_{21}}$ right is using $C_{\ell}^{E_{\mathrm{CMB}}B_{21}}$.}
%\label{fig:T_CMB_B_21}
%\end{figure}

%\begin{figure}
%    \centering    \includegraphics[width=0.5\linewidth]{figures/Sigma_val_window.pdf}
%    \caption{Cumulative error on $\alpha$ using the $C_{\ell}^{T_{21}B_{\mathrm{CMB}}}$ for the PUMA survey with the Dirac and Tophat window function}
%    \label{fig:placeholder}
%\end{figure}

\section{Conclusion and Outlook \label{eq:Conclu_Outlook}}
In this study, we gave constraints on birefringence angles $\alpha_\mathrm{CMB}$ and $\alpha_\mathrm{21cm}$ using futuristic 21cm and Planck-like CMB experiments.
%Constraints on birefringence angle have been $\alpha_\mathrm{CMB}$ 
%Previous studies of birefringence using CMB e constrained this parameter. Analyses of the Planck data release 4 report $\alpha = 0.30\degree\pm 0.11$ \cite{Diego-Palazuelos:2022mcp} while a recent study by Diego-Palazuelos and Komatsu combining data from the Atacama Cosmology Telescope (ACT) constrains the angle to $\alpha = 0.215\degree\pm0.074 \degree$, with a statistical significance of $2.9\sigma$ \cite{Diego-Palazuelos:2025dmh}. Complementary to CMB, the large-scale structures at late time can provide the additional information. Recent work by Yin et al. \cite{Yin:2024fez} demonstrates that galaxy surveys can provide the constraint of order $5\degree - 15 \degree$, and suggests that futuristic intensity mapping surveys like SKA CMB could offer additional enhancements. 
Our results indicate that if we cross-correlate the 21cm brightness temperature with the CMB, we can constrain the birefringence angle $\alpha_{\mathrm{CMB}} = 0.3\pm4.4\degree$ for the integrated PUMA experiment, though we don't get good constraints on $\alpha_\mathrm{21cm}$, where we get $\sigma_{\alpha_{21}} \approx {100}{\degree}$. %^These parameters have been constrained using six different correlations. Out of these 4 constrain $\alpha_{21cm}$ while two $\alpha_\mathrm{CMB}$. 
We found that the best constraints on these angles are given by cross correlations $C_\ell^{T_{21} B_\mathrm{CMB}}$ and $C_\ell^{T_{21} B_{21}}$ respectively. We further found that the constraints obtained on $\alpha_\mathrm{CMB}$ aren't as good as those obtained before using CMB-only limits. We believe that it is due to weaker cross correlation between 21cm and CMB signals (compared to what happens in CMB only case), which further happens due to large redshift separation. %Since we consider only after the reionization era, the overlap will be small. %The reason for not getting better constraints. {In our study we find that  due to the weak signal of 21cm E mode we are not getting better constraints from the $C_{\ell}^{EB}$ in comparison to $C_{\ell}^{TB}$}. {Can say something about the usage of bispectrum in our future endeavours.} {Future studies should extend beyond the power spectrum to investigate non-Gaussian signatures via the birefringence bispectrum. Extending recent CMB analyses \cite{Greco:2022ufo} to 21cm tomography could study the Lorentz violation and parity-breaking effects specifically during the Epoch of Reionization.}

\acknowledgments
Rahul Kothari acknowledges computing facilities availed through the IIT Mandi seed Grant IITM/SG/DIS-ROS-SPA/111. We are also truly thankful to Eiichiro Komatsu for the discussions.

\bibliographystyle{JHEP}
\bibliography{reference}
\end{document}